\documentclass[11pt]{article}
\usepackage[letterpaper, margin=1in]{geometry}
\usepackage[pdfusetitle]{hyperref}
\usepackage{setspace} 
\usepackage{authblk}
\usepackage{graphicx}
\usepackage[dvipsnames,svgnames,x11names]{xcolor}
\usepackage{doi}
\usepackage{cite}
\usepackage{booktabs}
\usepackage{amsmath,mathtools,amssymb}
\usepackage{cancel} 
\usepackage{textcomp,gensymb}

\usepackage{etoolbox}
\usepackage{multirow}
\usepackage{xcolor}
\usepackage{colortbl}

\makeatletter
\patchcmd{\env@cases}{1.2}{0.72}{}{}
\makeatother

\newcommand{\email}[1]{\footnote{\href{mailto:#1}{#1}}}

\hypersetup{
    colorlinks,
    linkcolor={red!50!black},
    citecolor={blue!50!black},
    urlcolor={blue!80!black}
}

\title{\Large\bf%
Charged Higgs Decay to Bottom and Charm Quarks from $Z_3$ - Flavored Two Higgs Doublet Models}
\author[1,2]{Alfredo Aranda\email{fefo@ucol.mx}}
\author[1,2,3]{J. Hern\'andez-S\'anchez\email{jaime.hernandez@correo.buap.mx}}
\author[4]{Andrea Montiel\email{mo337998@uaeh.edu.mx}}
\author[4]{R. Noriega-Papaqui\email{rnoriega@uaeh.edu.mx}}
\affil[1]{\textit{Facultad de Ciencias, Universidad de Colima, C.P. 28045, Colima, México}}
\affil[2]{\textit{Dual CP Institute of High Energy Physics, C.P. 28045, Colima, México}}
\affil[3]{\textit{Fac. de Cs. de la Electr\'onica, Benem\'erita Universidad Aut\'onoma de Puebla, Apdo. Postal 542, 72570 Puebla, Puebla, M\'exico}}
\affil[4]{\textit{\'Area Acad\'emica de Matem\'aticas y F\'{\i}sica, Universidad Aut\'onoma del Estado de Hidalgo, Carr. Pachuca-Tulancingo Km. 4.5, C.P. 42184, Pachuca, Hgo.}}

\begin{document}

\maketitle

\begin{abstract}
The phenomenology of a charged Higgs present in a model with two Higgs SU(2) doublets and a $Z_3$ flavor symmetry is analyzed. It is shown that it is possible to generate an enhancement of its flavor changing coupling to c and b quarks and also to reproduce the ATLAS excess associated to the process $H^\pm \to bc$ for a charge Higgs mass of $130$~GeV. Furthermore, by considering the possibility of a search at the future LHeC, the analysis suggests viability for its detection. 
\end{abstract}
\newpage

\section{Introduction}

It is difficult to conceive physics beyond the Standard Model (SM) where an extended scalar sector is not present. For any such nontrivial extension, the existence of additional neutral scalar fields as well as \emph{at least} an electrically charged scalar field are immediate physical consequences. These new states couple generally to SM fermions and lead to a rich yet strongly constrained phenomenology. The strong constraints imposed by the absence of flavor changing neutral currents (FCNC) imply a very restricted mixing in the Yukawa sector for the additional scalar states that must be either taken as a given or explained/induced by, for example, symmetry arguments. This has led to the study of the symmetries within the scalar sector itself, leading, for example in the case of two Higgs doublets models (THDM), to the well known results dubbed type I, II, and generalizations. For a comprehensive review see~\cite{Gunion:1989we,Branco:2011iw}. Recently, new so-called \emph{Goofy} symmetries have been found in the scalar sector of THDM that were previously unknown and are being studied ~\cite{Ferreira:2023dke,Trautner:2025yxz}. Once the Yukawa sector is also taken into consideration, strategies aimed at obtaining clues about the intriguing patterns and hierarchies found in fermion masses and mixing angles in the SM, have led to a wide spectrum of studies. Among those attempts, the interesting idea associated to flavor symmetries that differentiate among fermion generations has also led to a large variety of scenarios and models that typically involve extended \emph{flavored} scalar sectors~\footnote{There is a vast literature on these topics and any reference list will undoubtedly be incomplete. The authors apologize for all the omissions done by selecting a single review article.}. A nice review and pertinent references can be found in~\cite{Altarelli:2010gt}).

From the experimental perspective, having an electrically charged scalar state at a searchable energy scale provides a very interesting and important place to direct searches. The ATLAS collaboration has recently presented an analysis from a search using a proton-proton dataset from CERN's Large Hadron Collider (LHC) collisions at $\sqrt{s} = 13$~TeV, where upper limits at $95\%$ confidence-level of $0.15\%$ and $0.42 \%$ were obtained for the product of branching fractions $BR(t \to H^{\pm}b) \times BR(t \to cb)$ for $60$~GeV $\leq m_{H^{\pm}} \leq$ $160$~GeV~\cite{ATLAS:2023bzb}. Previous results by the CMS collaboration from a search of $H^{\pm} \to cb$ decays using collisions at $\sqrt{8}$~TeV reported $95\%$ confidence level limits on $BR(t \to H^{\pm}b)$ of $(0.8 - 0.5)\%$ (assuming $BR(H^{\pm} \to cb) = 1.0$) for $90$~GeV $\leq m_{H^{\pm}} \leq$ $150$~GeV~\cite{CMS:2018dzl}.

An interesting possibility one can entertain consists in exploring what the Large Hadron Electron Collider (LHeC)~\cite{LHeCStudyGroup:2012zhm}, being contemplated within the High-Luminosity LHC program, might be able to say about flavor changing processes induced by the charged Higgses of extended scalar sectors. This letter is motivated by such an idea and, in particular, considers a scenario where a discrete $Z_3$ \emph{flavor} symmetry helps differentiates among the charged Higgs couplings to fermion generations, potentially enhancing the decay channel $H^{+} \to \bar{b}c$. Section~\ref{sec:model} introduces the model followed by the numerical study of the charged Higgs boson in Section~\ref{sec:numerical}. The observation prospects for a light charged Higgs are then presented in Section~\ref{sec:prospects} and some final remarks summarizing the results in Section~\ref{sec:conclusion} conclude this letter. 

\section{The model} 
\label{sec:model}
The model contains the fermion content of the SM and two SU(2) doublets $H_i$, $i=1,2$, with the same hypercharge. There is a $Z_3$ flavor symmetry under which the fields transform as ${\cal F} \rightarrow  {\cal F}^{\prime} = \omega^{n_f} {\cal F}$, where $\omega \equiv \exp(2\pi i/3)$ and $|n_f| \in \{0,1,2\}$. We call $n_f$ the {\it charge} of the field ${\cal F}$ under $Z_3$: $[{\cal F}]=n_f$. For simplicity the flavor symmetry is assumed to act non-trivially only in the quark sector (we focus on the quark sector in this letter and leave the study of lepton mixing and neutrino masses for future investigation). An additional assumption of CP conservation is imposed on the scalar sector.

Denoting the SM three generations ($i=1,2,3$) left-handed quark and lepton doublets by $Q_i$ and $L_i$, the right-handed up-type quarks, down-type quarks, and charged leptons by $u_i$, $d_i$, and $\ell_i$, and finally the two Higgs doublets by $H_1$ and $H_2$, the $Z_3$ charge assignments are given by: $\left[ Q_1 \right] = 2$, $\left[ Q_2 \right] = \left[ Q_3 \right] = 1$, $\left[ u_1 \right] = 0$, $\left[ u_2 \right] = \left[ u_3 \right] = 1$, $\left[ d_1 \right] = 1$, $\left[ d_2 \right] = \left[ d_3 \right] = 2$, $\left[ L_i \right] = 0$, $\left[ \ell_i \right] = 0$, $\left[ H_1 \right] = 2$, and $\left[ H_2 \right] = 0$. These $Z_3$ charge assignments correspond to a case where the scalars couple to the up and down quark sectors in a \emph{flipped} way (see below). Note that $H_1$ does not participate in the lepton Yukawa sector.

The  SU(2)$_W\times$U(1)$_Y \times Z_3$ invariant scalar potential can be written as
\begin{eqnarray} \label{potential} \nonumber
V(H_1,H_2) &=& \mu_{1}^2 H_1^{\dagger}H_1 + \mu_{2}^2 H_2^{\dagger}H_2  + \tilde{\mu}^2 \left( H_1^{\dagger}H_2 + h.c. \right) + \lambda_{1} H_1^{\dagger}  H_1H_1^{\dagger}H_1 \\ 
&+& \lambda_{2} H_2^{\dagger}H_2H_2^{\dagger}H_2
+ \lambda_{3} H_1^{\dagger}H_1H_2^{\dagger}H_2 + \lambda_{4} H_1^{\dagger}H_2H_2^{\dagger}H_1 \ ,
\end{eqnarray}
where a soft-breaking term has been included in order to avoid the presence of an extra Goldstone Boson. Each scalar field acquires a vacuum expectation value (vev) $v_i$ and is generally expressed as
\begin{eqnarray}
\label{scalars}
H_i = 
\left(  
	\begin{array}{c}
	H_i^+ \\
	\frac{1}{\sqrt{2}}(h_i + v_i + i A_i) 
	\end{array}
\right) \ ,
\end{eqnarray}
where $v_i$ denotes the vev of $H_i$. Using this into Eq.~\eqref{potential} we obtain the following squared scalar mass matrices:
\begin{eqnarray} \label{scalarmassmatrix}
{\cal M}^2_S = 
\left(
	\begin{array}{cc}
	2v_1^2\lambda_1 -\frac{v_2}{v_1}\tilde{\mu}^2 & v_1v_2 (\lambda_3 + \lambda_4) + \tilde{\mu}^2 \\
	v_1v_2 (\lambda_3 + \lambda_4) + \tilde{\mu}^2  & 2v_2^2 \lambda_2 -\frac{v_1}{v_2}\tilde{\mu}^2
	\end{array}
\right) ,
\end{eqnarray}
\begin{eqnarray}\label{pscalarmassmatrix}
{\cal M}^2_{PS} = 
\tilde{\mu}^2
\left(
	\begin{array}{cc}
	-\frac{v_2}{v_1} & 1 \\
	1 & -\frac{v_1}{v_2}
	\end{array}
\right) \longrightarrow 
{\cal M}^2_{PSD} = 
\left(
	\begin{array}{cc}
	0 & 0 \\
	0 & -\tilde{\mu}^2\frac{(v_1^2+v_2^2)}{v_1v_2}
	\end{array}
\right),
\end{eqnarray}
and 
\begin{eqnarray}\label{chargedmassmatrix}
{\cal M}^2_{C} = 
\left(
	\begin{array}{cc}
	-\frac{v_2}{2v_1}(2\tilde{\mu}^2+v_1v_2\lambda_4) & \tilde{\mu}^2+ \frac{v_1v_2\lambda_4}{2}\\
	 \tilde{\mu}^2+ \frac{v_1v_2\lambda_4}{2} & -\frac{v_1}{2v_2}(2\tilde{\mu}^2+v_1v_2\lambda_4)
	\end{array}
\right) \longrightarrow 
{\cal M}^2_{CD} = 
\left(
	\begin{array}{cc}
	0 & 0 \\
	0 & -(2\tilde{\mu}^2+v_1v_2\lambda_4)\frac{(v_1^2+v_2^2)}{2v_1v_2}
	\end{array}
\right)\ .
\end{eqnarray}
The Yukawa sector is obtained from the following expression
\begin{eqnarray}\label{yukawa}
-{\cal L}_Y = {\cal Y}^{ua}_{ij} \overline{Q}_i \widetilde{H}_a u_j + {\cal Y}^{da}_{ij} \overline{Q}_i H_a d_j + h.c. \ ,
\end{eqnarray}
where $a=1,2$, $i,j=1,2,3$, and $\widetilde{H}_i \equiv i\sigma_2H_i^*$. After the spontaneous breaking of the gauge symmetry, the quark mass matrices take the form
\begin{eqnarray}\label{finalUyukawas-case1}
{\cal M}^u \sim
\left( \begin{array}{ccc}
0 & v_1 & v_1 \\
v_1 & v_2 & v_2 \\
v_1 & v_2 & v_2
\end{array}\right), \qquad
{\cal M}^d \sim
\left( \begin{array}{ccc}
0 & v_2 & v_2 \\
v_2 & v_1 & v_1 \\
v_2 & v_1 & v_1
\end{array}\right).
\label{Mq}
\end{eqnarray}
where only the vev dependence has been included and the arbitrary coefficients for each entry in the matrices have been omitted in order to note the peculiar {\it flipping} of the vev structure in the mass matrices, i.e. each entry of these matrices has an arbitrary coefficient that must be included in order to perform the numerical analysis. 

\section{Parametrization and Numerical analysis}
\label{sec:numerical}
\subsection{Yukawa parameters and couplings}
After electroweak symmetry breaking (EWSB), the up-type and down-type quark mass matrices become
\begin{eqnarray}
    {\cal M}^q=\frac{1}{\sqrt{2}} (v_1 {\cal Y}^{q1}+v_2 {\cal Y}^{q2}), \ \ \ q=u,d \ ;
\end{eqnarray}
where, given the $Z_3$ charge assignments for fermions and scalars, the Yuakawa parameter matrices ${\cal Y}^{qi}$ have the textures:
\begin{eqnarray}\label{Y-matrices}
v_1 {\cal Y}^{u1} =
v_1 \left( \begin{array}{ccc}
0 &  {\cal Y}^{u1}_{12} &  {\cal Y}^{u1}_{13} \\
{\cal Y}^{u1}_{21}  & 0 & 0 \\
{\cal Y}^{u1}_{31}  &0 & 0
\end{array}\right), \qquad
v_2 {\cal Y}^{u2} =
v_2 \left( \begin{array}{ccc}
0 &  0 &  0 \\
0  & {\cal Y}^{u2}_{22} & {\cal Y}^{u2}_{23} \\
0  & {\cal Y}^{u2}_{32} & {\cal Y}^{u2}_{33}
\end{array}\right), \,  \, \\
v_1 {\cal Y}^{d1} =
v_1 \left( \begin{array}{ccc}
0 &  0 &  0 \\
0  & {\cal Y}^{d1}_{22} & {\cal Y}^{d1}_{23} \\
0  & {\cal Y}^{d1}_{32} & {\cal Y}^{d1}_{33}
\end{array}\right), \qquad
v_2 {\cal Y}^{d2} =
v_2 \left( \begin{array}{ccc}
0 &  {\cal Y}^{d2}_{12} &  {\cal Y}^{d2}_{13} \\
{\cal Y}^{d2}_{21}  & 0 & 0 \\
{\cal Y}^{d2}_{31}  &0 & 0
\end{array}\right). \,  
\label{Mqd}
\end{eqnarray}
For each quark family, the above matrices ${\cal Y}^{qi}$ are not aligned with each other and not necessarily diagonalized simultaneously with the mass matrix. From now on, ad additional assumption is made that these matrices are Hermitian. ${\cal M}^{qi}$ are diagonalized by the unitary transformation:
\begin{equation}
    \hat{{\cal M}^q}={\cal U}_q^{\dagger} {\cal M}^q {\cal U}_q = \text{Diag}(\lambda_1^q, \lambda_2^q, \lambda_3^q)=\frac{1}{\sqrt{2}} (v_1 \hat{\cal{Y}}^{q1}+v_2 \hat{\cal{Y}}^{q2}), \ \ \ q = u, d \ ;
    \label{Mqdiag}
\end{equation}
where the matrix ${\cal U}_q$ is constructed as the product of the two matrices ${\cal P}_q$ and ${\cal O}_q$. The matrix ${\cal P}_q$ removes the phases, while the matrix ${\cal O}_q$ contains the normalized eigenvectors of the phases-free matrix. The eigenvalues $\lambda_i^q$ define the absolute values  $|m_{qi}|$ that correspond to the masses of the quarks $q_i$. Finally, $\hat{\cal{Y}}^{qi} = {\cal U}_q^{\dagger}\, {\cal{Y}}^{qi}\, {\cal U}_q$ ($n=1,2$). 
Defining $v_1= v \cos{\beta}$ and $v_2= v \sin{\beta}$ (with $v=246$ GeV), the mass matrices ${\cal M}^u$ and ${\cal M}^d$ become:
\begin{equation}
\label{Ms}
{\cal M}^u  = v_1
\begin{pmatrix}
 0 & a_{12} & a_{13}\\
 a^*_{12} & a_{22} \tan{\beta} & a_{23} \tan{\beta}\\
 a^*_{13}& a^*_{23} \tan{\beta} & a_{33} \tan{\beta}\\
\end{pmatrix}, \qquad
{\cal M}^d  = v_1
\begin{pmatrix}
 0 & b_{12} \tan{\beta} & b_{13}\tan{\beta}\\
 b^*_{12} \tan{\beta}& b_{22}  & b_{23} \\
 b^*_{13} \tan{\beta}& b^*_{23}  & b_{33} \\
\end{pmatrix} \ ,
\end{equation}
where $a_{ij}$ and $b_{ij}$ denote arbitrary parameters whose values must be determined and reproduce the experimental values of the quark masses and mixing angles in CKM matrix \(V_{\mathrm{CKM}}\)~\footnote{Note the \emph{flipped} nature of the mass matrices in terms of $\tan\beta$, which acts as a weight function and thereby introduces a distinct flavor dynamics.}. It is important to note that these parameters are not all independent, as they are constrained by the following relations:
\begin{align}
\textrm{Tr}\left[{\cal M}^{q\dagger} {\cal M}^q \right] & =  (\lambda_1^q)^2+(\lambda_2^q)^2+(\lambda_3^q)^2,\\
\textrm{Det}\left[{\cal M}^{q\dagger} {\cal M}^q\right] & =   (\lambda_1^q)^2(\lambda_2^q)^2(\lambda_3^q)^2,\\
\frac{1}{2}\left[\textrm{Tr}^2\left[{\cal M}^{q\dagger} {\cal M}^q \right]- \left( \textrm{Tr}(\left[{\cal M}^{q\dagger} {\cal M}^q \right] \right)^2 \right] & =  (\lambda_1^q)^2(\lambda_2^q)^2+(\lambda_1^q)^2(\lambda_3^q)^2 
 +(\lambda_2^q)^2(\lambda_3^q)^2.
\end{align}

\begin{figure}[ht]
    \centering
    \includegraphics[width=0.85\linewidth]{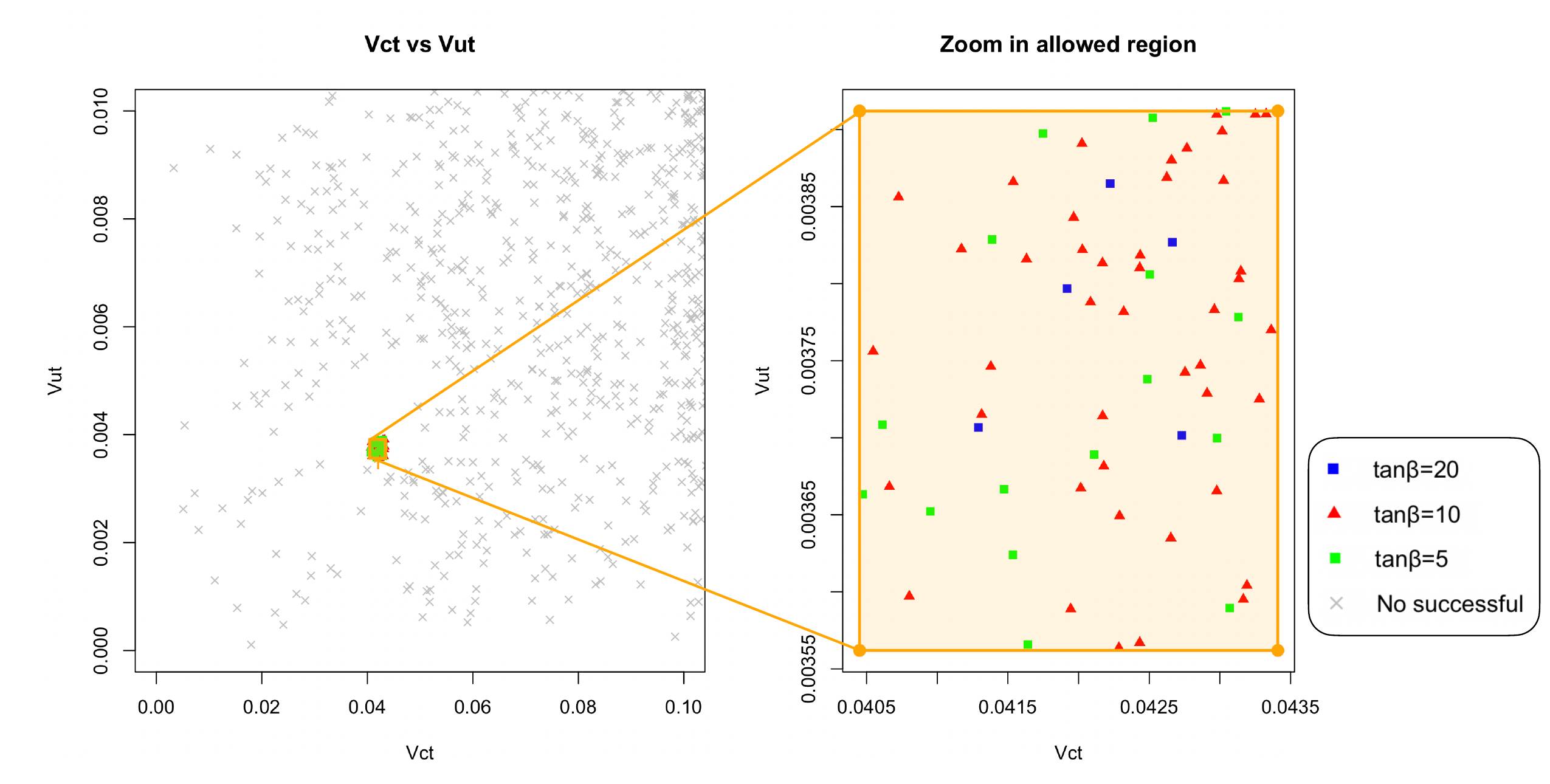}
    \caption{Left: Values of $V_{ct}$ and $V_{ut}$ for sets of parameters $\vec{s}=(a_{ij},b_{ij})$ that reproduce quark masses values and $V_{CKM}$ entries (other than $V_{ct}$ and $V_{ut}$) for three values of $\tan\beta$. The small rectangular area corresponds to the experimentally allowed values of $V_{ct}$ and $V_{ut}$ where only $125$ sets remain ($25$ for each value of $\tan \beta$). Right: A zoom of the region containing sets consistent with all quark masses and $V_{CKM}$ entries.}
    \label{Fig:Vckm:ver}
\end{figure}

The above equations are highly nonlinear and the parameters are determined numerically. To this end, the well-established bio-inspired optimization method Particle Swarm Optimization (PSO) algorithm has been employed~\footnote{The code implemented in this work can be find in: \url{https://github.com/MontielAnn/Z3-chargedH-bc.git}}~\cite{CuevasPSO,WileyPSO}. Fixing $\tan\beta$ to the values $1, 2, 5, 10, 20$ the PSO then finds sets of consistent parameters $\vec{s}=(a_{ij},b_{ij})$. All CKM matrix elements were computed and verified numerically and the most restrictive ones correspond to $V_{ct}$ and $V_{ut}$. Figure~\ref{Fig:Vckm:ver} shows these results in the $V_{ct} - V_{ut}$ plane: the left panel contains the values obtained for $V_{ct}$ and $V_{ut}$ consistent with quark masses and $V_{CKM}$ entries (other than $V_{ct}$ and $V_{ut}$), and the small rectangular region allowed by the experimental values for $V_{ct}$ and $V_{ut}$. The right panel zooms in onto the allowed region for both $V_{ct}$ and $V_{ut}$. 

In order to further constraint the model parameters, the strategy presented in~\cite{Hernandez-Sanchez:2012vxa} was followed for the Yukawa sector. Expressing the two scalar fields by $h^0$ and $H^0$, the pseudoscalar by $A^0$, and the charged one by $H^{\pm}$, the generic lagrangian describing the interactions of the charged and neutral Higgs bosons can be written as:
\begin{eqnarray} \nonumber
     -{\cal L}_{\phi f f} &=& \frac{\bar{U}_L}{\sqrt{2}} \bigg[ (-s_\alpha {\cal \hat{Y}}^{u1}+ c_\alpha {\cal \hat{Y}}^{u2}) h^0+ (c_\alpha {\cal \hat{Y}}^{u1}+ s_\alpha {\cal \hat{Y}}^{u2}) H^0+i (s_\beta {\cal \hat{Y}}^{u1}- c_\beta {\cal \hat{Y}}^{u2}) A^0 \bigg] U_R \\ \nonumber
     &+& \frac{\bar{D}_L}{\sqrt{2}} \bigg[ (-s_\alpha {\cal \hat{Y}}^{d1}+ c_\alpha {\cal \hat{Y}}^{d2}) h^0+ (c_\alpha {\cal \hat{Y}}^{d1}+ s_\alpha {\cal \hat{Y}}^{d2}) H^0-i (s_\beta {\cal \hat{Y}}^{d1}- c_\beta {\cal \hat{Y}}^{d2}) A^0 \bigg] D_R  \\ \nonumber
     &+&\bar{U}_R \left(s_\beta {\cal \hat{Y}}^{u1}- c_\beta {\cal \hat{Y}}^{u2}\right)\cdot \left(V_{CKM}\right) D_L H^+  - \bar{U}_L  (V_{CKM}) \cdot \left(s_\beta {\cal \hat{Y}}^{d1}- c_\beta {\cal \hat{Y}}^{d2}\right) D_L H^+ \\  
     &+& \frac{\sqrt{2}}{v} \cot{\beta} H^+ \bar{N}_L {\cal \hat{M}}^l E_R + \frac{1}{v \sin{\beta} } \left(H^0 s_\alpha + h^0 c_\alpha +i A^0 c_\beta \right) E_L {\cal \hat{M}}^l L_R+ h.c. \ ,
\end{eqnarray}
with $U_{L,R} \equiv (u_{L,R}, c_{L,R}, t_{L,R})$, $D_{L,R} \equiv (d_{L,R}, s_{L,R}, b_{L,R})$, $N_{L} \equiv (\nu_{eL}, \nu_{\mu L}, \nu_{\tau L})$, $E_{L,R} \equiv (e_{L,R}, \mu_{L,R}, \tau_{L,R})$, and where $c_\alpha = \cos \alpha$, $s_\alpha = \sin \alpha$, $c_\beta = \cos \beta$ and $s_\beta =\sin \beta$. It is relevant to stress that the Yukawa matrices are linearly independent and not aligned with each other. In order to explore possible enhancements for the  $H^\pm cb$ coupling, it becomes useful to invoke the known THDM configuration, and following~\cite{Hernandez-Sanchez:2012vxa}, one can define two configurations for the model of this letter (denoted from now on by THDMZ3):
\begin{itemize}
    \item THDMZ3 Type-A
    \begin{eqnarray}
        {\cal \hat{Y}}^{u2} =\frac{\sqrt{2}}{v\sin \beta} {\cal \hat{M}}^u - \cot{\beta} {\cal \hat{Y}}^{u1},  \ \ \  {\cal \hat{Y}}^{d2} =\frac{\sqrt{2}}{v\sin \beta} {\cal \hat{M}}^d - \cot{\beta}  {\cal \hat{Y}}^{d1}   
    \end{eqnarray}
     \item THDMZ3 Type-B
     \begin{eqnarray}
        {\cal \hat{Y}}^{u2} =\frac{\sqrt{2}}{v\sin \beta} {\cal \hat{M}}^u - \cot{\beta}  {\cal \hat{Y}}^{u1},  \ \ \  {\cal \hat{Y}}^{d1} =\frac{\sqrt{2}}{v\cos \beta} {\cal \hat{M}}^d - \tan{\beta}  {\cal \hat{Y}}^{d2}
         \end{eqnarray}
\end{itemize}
Then the generic interactions of fermions with scalars can be rephrased as:  
\begin{eqnarray} \nonumber
   {\cal L}_{H^\pm u_i d_j} & = & -\frac{1}{v}\bar{f_i}
   \left(h_{ij}^f h^0 +H_{ij}^f H^0 - i A_{ij}^f \gamma_5 A^0\right) f_j \\
   &-&\frac{\sqrt{2}}{v} \left[
   \bar{U}_i (m_{d_j} X_{ij} P_R + m_{u_i} Y_{ij} P_L)D_j + Z \bar{\nu}_L l_R \right] H^+
   + h.c.
\end{eqnarray}
where $ Z=-\cot \beta$ and the couplings $X_{ij}$, $Y_{ij}$,  $h_{ij}$, $H_{ij}$, $A_{ij}$ are given as a functions of the mixing angles $\alpha$ and $\beta$ by:
\begin{itemize}
    \item For the THDMZ3 Type-A:
    \begin{eqnarray}
    X_{ij} &=& (V_{\small CKM})_{il} \bigg( X \delta_{lj} -\frac{v}{\sqrt{2}} \frac{f(X)}{ md_j  }  {\cal \hat{Y}}_{lj}^{d1} \bigg), \, \,\,\,\,     Y_{ij}= \bigg( Y \delta_{il} -\frac{v}{\sqrt{2}}\frac{f(Y)}{ mu_i }  {\cal \hat{Y}}_{il}^{u1} \bigg) (V_{CKM})_{lj},  \\
    h_{ij}^f  &=&  \bigg(\frac{c_\alpha}{s_\beta} {\cal \hat{M}}_{ij}^f- \frac{v}{\sqrt{2}} \frac{c_{\beta-\alpha}}{s_\beta} {\cal \hat{Y}}_{ij}^{f1}\bigg), \,\,\ \,\,\,\,\,\,\,\,\,\,    \,\,\,\,\, \,\,\,\,
    H_{ij}^f  = \bigg(\frac{s_\alpha}{s_\beta} {\cal \hat{M}}_{ij}^f+ \frac{v}{\sqrt{2}} \frac{s_{\beta-\alpha}}{s_\beta} {\cal \hat{Y}}_{ij}^{f1}\bigg), \, \,\,\,\,\,\\
   A_{ij}^u&=& \bigg( Y {\cal \hat{M}}_{ij}^u -\frac{v}{\sqrt{2}}f(Y)  Y^{u 1}_{ij} \bigg) , \,\,\,\,\,\,  \,\,\,\,\,\,\,\,   \,\,\,\,\,\,\,\,\,\,\,\,\,      A_{ij}^d= \bigg(- X {\cal \hat{M}}_{ij}^d +\frac{v}{\sqrt{2}}f(X)  Y^{d 1}_{ij} \bigg),  
   \label{couplings-typeI}
\end{eqnarray}
with $X=-Y=Z$ and $f(x)=\sqrt{1+x^2}$.
 \item For the THDMZ3 Type-B ($X=\tan \beta$, $X=1/Y$):
 \begin{eqnarray}
    X_{ij} &=& (V_{CKM})_{il} \bigg( X \delta_{lj} -\frac{v}{\sqrt{2}} \frac{f(X)}{ md_j  }  {\cal \hat{Y}}_{lj}^{d2} \bigg), \, \,\,\,\,     Y_{ij} = \bigg( Y \delta_{il} -\frac{v}{\sqrt{2}}\frac{f(Y)}{ mu_i }  {\cal \hat{Y}}_{il}^{u1} \bigg) (V_{CKM})_{lj},  \\
    h_{ij}^u  &=&  \bigg(\frac{c_\alpha}{s_\beta} {\cal \hat{M}}_{ij}^u- \frac{v}{\sqrt{2}} \frac{c_{\beta-\alpha}}{s_\beta} {\cal \hat{Y}}_{ij}^{u1}\bigg),   \,\,\,\,\,\,\,\,\,\, \,\,\,\,\,\,\,\,\,\, \,\,\,\,
     h_{ij}^d  =  \bigg(-\frac{s_\alpha}{c_\beta} {\cal \hat{M}}_{ij}^d+ \frac{v}{\sqrt{2}} \frac{c_{\beta-\alpha}}{c_\beta} {\cal \hat{Y}}_{ij}^{d2}\bigg), \\
    H_{ij}^u  &=& \bigg(\frac{s_\alpha}{s_\beta} {\cal \hat{M}}_{ij}^u+ \frac{v}{\sqrt{2}} \frac{s_{\beta-\alpha}}{s_\beta} {\cal \hat{Y}}_{ij}^{u1}\bigg), \, \, \,\,\,\,\,\,  \,\,\,\,\,\,\,\,\,\,    \,\,\,
    H_{ij}^d \, =  \bigg(\frac{c_\alpha}{c_\beta} {\cal \hat{M}}_{ij}^d- \frac{v}{\sqrt{2}} \frac{s_{\beta-\alpha}}{c_\beta} {\cal \hat{Y}}_{ij}^{d2}\bigg),\\
   A_{ij}^u &=& \bigg( Y {\cal \hat{M}}_{ij}^u -\frac{v}{\sqrt{2}}f(Y)  Y^{u 1}_{ij} \bigg) , \,\,\,\,\,\,  \,\,\,\,\,\,\,\,   \,\,\,\,\,\,\,\,\,\, \,\,      A_{ij}^d = \bigg( X {\cal \hat{M}}_{ij}^d -\frac{v}{\sqrt{2}}f(X)  Y^{d 2}_{ij} \bigg). 
   \label{couplings-typeII}
\end{eqnarray}
\end{itemize}
Since the scalar potential is a particular case of the general THDM, it is possible to obtain the parameter space allowed by electroweak precision observables (for instance the oblique parameters), as well as by theoretical constraints such as vacuum stability, unitarity and perturbativity. Taking this into account and considering a scenario where the charged Higgs could be light, the following   parameter space is selected: $m_{h^0}=125$~GeV (corresponding to the SM-like Higgs boson), $m_{A^0}>m_{H^\pm}$ (avoiding the channel decay $H^\pm \to AW^{\pm *}$\cite{Akeroyd:2012yg, Akeroyd:2022ouy}), $160$~GeV$ \leq m_{H^0} \leq 260$~GeV, and $80$~GeV $\leq m_{H^\pm} \leq 170$~GeV with $\cos{(\beta-\alpha)} \leq 0.1$.

\subsection{Flavor and Higgs physyics constraints}
Following the analyses in~\cite{Hernandez-Sanchez:2012vxa,Crivellin:2013wna,Trott:2010iz}, the parameter space of the model is constrained by considering all relevant experimental bounds on flavor physics, i.e. 
bounds coming from leptonic and semileptonic decays mesons, $b\to s \gamma$, $B_0-\overline{B}_0$ mixing (compatible with $K_0-\overline{K}_0$ mixing), and the neutron's electric dipole moment ($d_n$) (see also the analysis in~\cite{Akeroyd:2022ouy}). The constraints coming from $B$ and $d_n$ physics are the strongest. 

\begin{center}
    \begin{figure}[ht]
    \centering
    \includegraphics[width=0.75\linewidth]{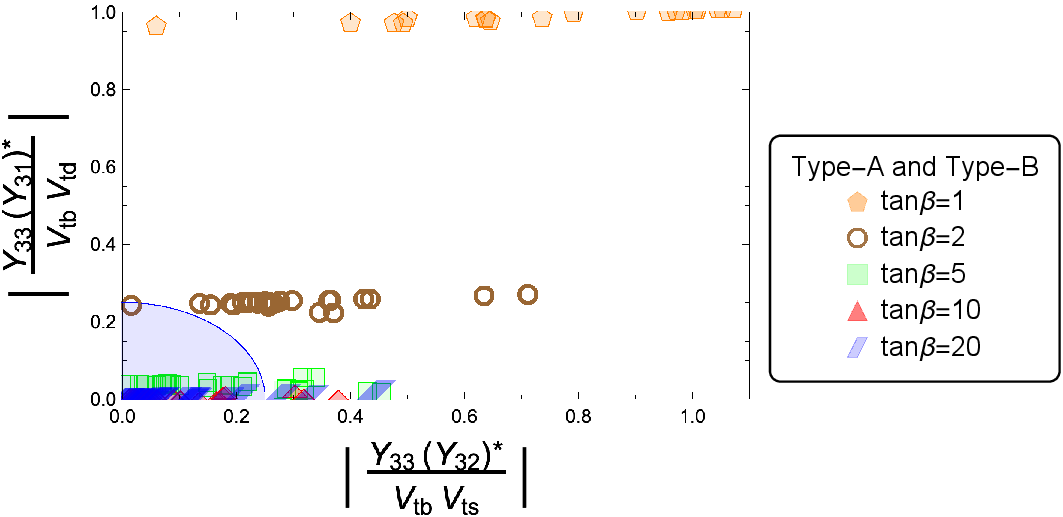}
    \caption{We apply, to the survivor parameter space, one of the strongest experimental limits at low energies $b \to s \gamma$  and $B_0-\overline{B}_0$ mixing. The shaded region is the allowed by both constraints.}
    \label{fig:BB}
\end{figure}
\end{center}

To perform the analysis of the model in this letter, the following combination of $b\to s \gamma$ and $B_0-\overline{B}_0$ mixing (for $m_{H^\pm } = 100 $~GeV) is utilized:
\begin{eqnarray}
    \bigg| \frac{Y_{33} Y_{3,i}^*}{V_{tb} V_{t d_i}}  \bigg| \leq 0.25, \,\,\,\,\,\,\, i=1,2.
\label{BB}
\end{eqnarray}
 This filter is applied to the set of points compatible with $V_{CKM}$ and mass matrices obtained above. As can be seen in Figure~\ref{fig:BB}, the case with $\tan \beta =1$ does not survive this bound and, for $\tan \beta =2$, only one point lies within the allowed region. For $\tan \beta =5$, $10$, and $20$, several points pass this filter. As a second step, those points in the allowed region are then examined under the constraints imposed by the combination of $b\to s\gamma $ and the electric dipole moment of the neutron ($d_n$) bound, given respectively by: 
\begin{eqnarray}
-1.1 \leq Re\bigg( \frac{X_{33} Y_{3,2}^*}{V_{tb} V_{t s}} \bigg)  \leq 0.7, \,\,\,\,\,\,\,\,\,
    \bigg| Im\bigg( \frac{X_{33} Y_{3,2}^*}{V_{tb} V_{t s}} \bigg) \bigg|  \leq 0.1.
\label{Bdn}
\end{eqnarray}

\begin{figure}[ht]
    \centering
    \includegraphics[width=0.6\linewidth]{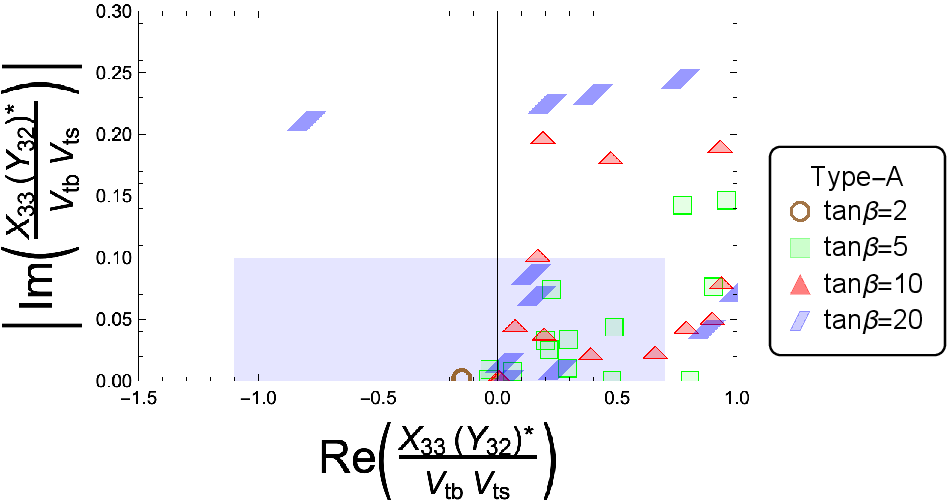}
    \vspace{1cm}
    \\
    \includegraphics[width=0.6\linewidth]{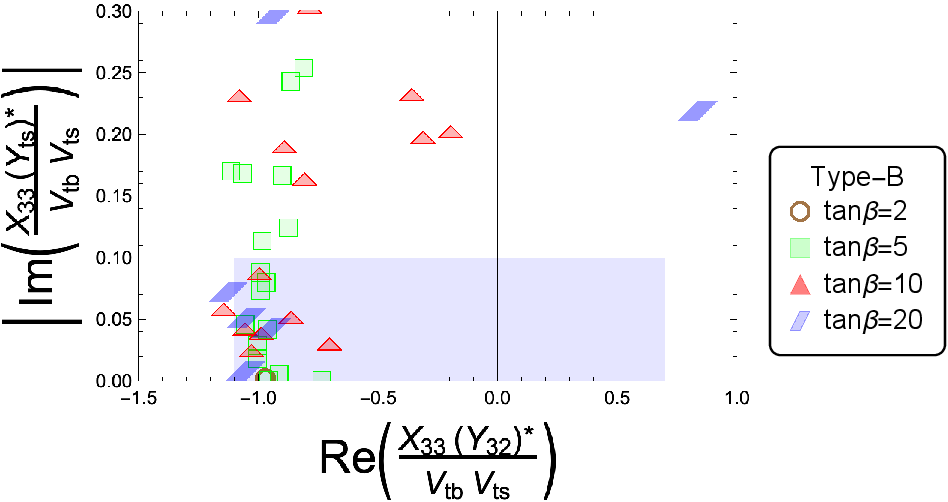}
    \caption{Considering the points that survive in Figure~\ref{fig:BB}, we now apply the strongest constraints that coming from $b \to s \gamma$ and the electric dipole moment of the neutron. The shaded region is the allowed by both constraints.}
    \label{fig:Bdn}
\end{figure}
Figure~\ref{fig:Bdn} shows that, albeit the $d_n$-bound is quite strong, several points are in agreement with it. The next step is to impose Higgs physics constraints on these points. To do so, the $\kappa$-formalism or $\kappa$-framework~\cite{ParticleDataGroup:2024cfk,LHCHiggsCrossSectionWorkingGroup:2012nn} is adopted, which rescales the SM Higgs boson couplings ($\kappa_i = g_{h^0 jj}/g_{h_{SM}^0 jj}$) provided the same Lorentz structure is maintained. This parametrization is associated with experimental data of the two-body decay channels  (or production) of the Higgs boson $h^0$ by $\kappa_i^2= \Gamma_i/\Gamma_i^{SM}$,  ($\kappa_i^2= \sigma_i/ \sigma_i^{SM}$). Taking all the points that pass the constraints of Figure~\ref{fig:Bdn} and whose couplings~\eqref{couplings-typeI}-\eqref{couplings-typeII} satisfy the most recent results for ATLAS and CMS Run 2~\cite{ParticleDataGroup:2024cfk}, the constraints shown in Table~\ref{tab:higgs} are obtained. Once these constraints are added, the allowed points in Figure~\ref{fig:Bdn} are Benchmark Points (BP) candidates  of the model. 

Summarizing: the benchmark points candidates selected are the ones with $m_{h^0}=125$ GeV, $m_{A^0}>m_{H^\pm}$, 160 GeV$ \leq m_{H^0} \leq 260$ GeV, and 80 GeV $\leq m_{H^\pm} \leq 160$ GeV with $\cos{(\beta-\alpha)} \leq 0.008$ for $\tan \beta=2$, $5$, $10$, and $20$. 
   \begin{center}
   \begin{table}[ht]
       \begin{tabular}{|c|c|c|c|}
       \hline 
       \cellcolor{gray!25} \ $\kappa_i$ \ & \ \cellcolor{gray!25} CMS or ATLAS Run 2 \ & \cellcolor{gray!25} THDMZ3 Type-A & \cellcolor{gray!25} 2HDMZ3 Type B
       \\
       \hline \hline
      $\kappa_\tau  $   &  $0.91\pm 0.07$ CMS   & $ \cot{\beta} \leq 0.91 $ & $1.02 \leq \tan \beta$  \\ \hline
      $\kappa_b$  & $0.98^{+0.13}_{-0.12}$ CMS & $|\cos{(\beta-\alpha)}| \leq 0.01$ & $|\cos{(\beta-\alpha)}| \leq 0.008$  \\  \hline
      $\kappa_t$ & $0.99\pm 0.09$  ATLAS & $|\cos{(\beta-\alpha)}| \leq 0.01$ & $|\cos{(\beta-\alpha)}| \leq 0.01$ \\ \hline
$k_\gamma$ & 0.97 $\pm 0.06$ ATLAS & \ \  80 GeV $\leq m_{H^\pm}\leq 170$ GeV \ \   & \ \  80 GeV $\leq m_{H^\pm}\leq 160$ GeV  \ \  \\
\hline
       \end{tabular}
       \caption{Constraints on the allowed points from Figure~\ref{fig:Bdn} (for $\tan \beta = 2$, $5$, $10$, $20$). }
              \label{tab:higgs}
   \end{table}
  \end{center}

Note that for neutral Higgs bosons the FCNC constraints given in~\cite{Chiang:2017etj} have been implemented, where the parameter that introduces FCNF is given by $\bar{Y}_{ij}=\sqrt{2 m_i m_j}/v$, with
\begin{eqnarray}
    \bar{Y}_{ij}^f\sim h_{ij}^f/v  \sim \frac{z}{v}  \frac{\sqrt{1+ \tan\beta^2}}{\tan\beta} \ ,
\end{eqnarray}
with $z/v \leq 10 ^{-4}$ for $h_{SM}=h^0$, in agreement with the current experimental constraint. Additionally, following the analysis in~\cite{Hernandez-Sanchez:2012vxa},  we consider the strongly restrictive constraints arising from $B_s \to \mu^+ \mu^-$, for both $A^0$ and $H^0$, obtaining $z/v \leq 10 ^{-3}$, also in agreement with current bounds.

\subsection{Charged Higgs Boson phenomenology}
   In order to study the dominant production of a light charged Higgs boson ($H^\pm$) and to compare the results of the model to experimental data from the LHC, the light $H^\pm$ decays must be analyzed. In particular, the expressions for the partial widths to fermions are reduced to:
   \begin{eqnarray}
       \Gamma (H^\pm \to u_i b_j) &=& \frac{3G_F m_{H^\pm} (m_{dj}^2 |X_{ij}|^2+ m_{ui}^2 |Y_{ij}|^2)}{4 \pi \sqrt{2}}  \ , \\
        \Gamma (H^\pm \to \ell \nu_\ell ) &=& \frac{G_F m_{H^\pm} (m_{\ell}^2 |\cot \beta|^2)}{4 \pi \sqrt{2}} \ ,
   \end{eqnarray}
   where the running quark masses are evaluated at the scale ($Q=m_{H^\pm}$), and the QCD vertex corrections ($1+ 17 \alpha_s^2/(3\pi))$ have been considered. Using the benchmark points of the model leads to the dominant decay channel $H^\pm \to cb$ due to the hierarchies $X_{23} m_b> X_{22} m_s>Z$ and  $X_{23} m_b > Y_{22} m_c$ for THDMZ3 Type-A for $\tan \beta=5$, 10; while for THDMZ3 Type-B   $X_{23}> X_{22} >Z$ and  $Y_{22}> Y_{23}$ when $tan \beta = 5$, 10, 20. Note that both types are consequence of the textures of the quarks matrices - or $Z_3$ flavor symmetry,  whose effect is observed in the parameter $\tan \beta$. Also, the fact that $ms (Q=2\ \rm{GeV})= 99$~MeV and $ms (Q=m_{H^\pm})= 55$~MeV (with $m_{H^\pm}=130$~GeV) is relevant for the calculations of the branching ratios of the charged Higgs  \cite{Akeroyd:2012yg,Hernandez-Sanchez:2012vxa,Akeroyd:2022ouy}. 
\begin{center}
    \begin{figure}[ht]
    \centering
    \includegraphics[width=0.6\linewidth]{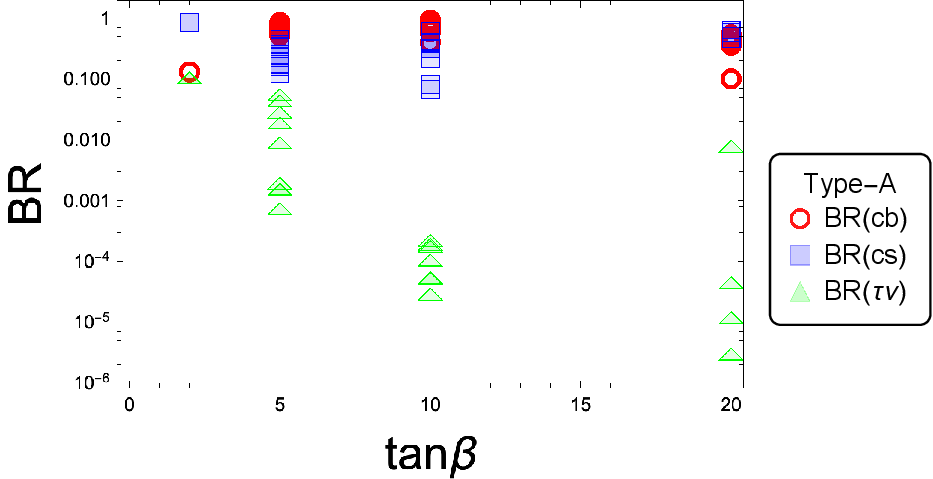}
    \vspace{1cm}
    \\
    \includegraphics[width=0.6\linewidth]{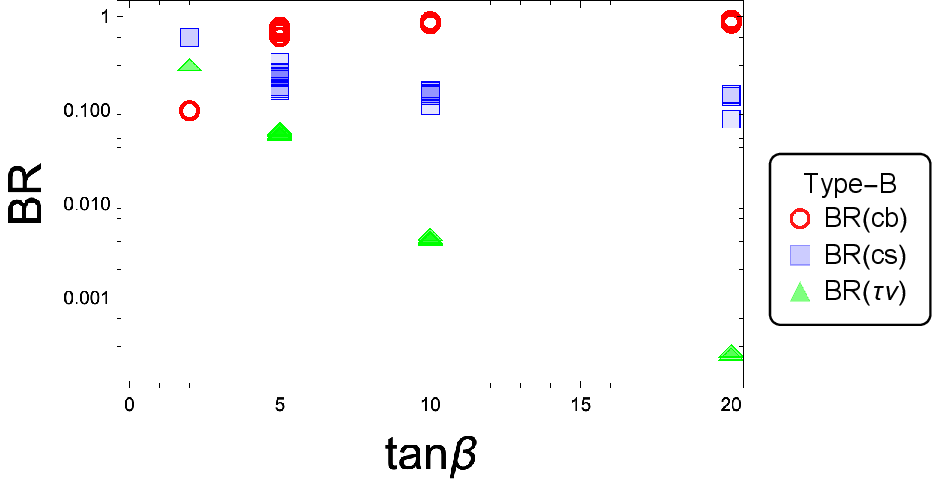}
    \caption{$BR( H^\pm \to c b,cs,\tau \nu)$ for the benchmark points of the model. The effect of the $Z_3$ flavor symmetry on the quark sector can be seen in the enhancement of the channel decay $H^\pm \to c b$, which is dominant with $BR \sim 70 \%$ for $\tan \beta=5$, $10$, $20$. The upper panel shows the THDMZ3 Type-A case while the THDMZ3 Type-B case in shown on the bottom panel.}
    \label{fig:BRs}
    \end{figure}
\end{center}
Figure~\ref{fig:BRs} shows the branching ratios $BR(H^\pm \to cb, cs, \tau \nu)$. It can be seen that the dominant decay is $H^\pm \to cb$, for both THDMZ3 Type-A (THDMZ3 type -B), with $\tan \beta = 5$, $10$ ($\tan \beta = 5$, $10$, and  $20$ ),  and the $BR(cb) > 70 \% $ ($BR(cb)=60-90 \%$ ) can be reached. 
This analysis is important for the dominant production mechanism in the LHC, which for a light $H^\pm$, corresponds to $pp \to t\bar{t}$ followed by the decay $t \to H^\pm b$, which then allows $H^\pm $ to decay into one of these modes: $cb, cs, cs+cb,\tau \nu$. Thus, the product $BR(t \to H^\pm b) \times BR(H^\pm \to cb/ cs/ cb+cs/\tau \nu)$ is important for the statistics analysis of the final states of the process, and in this model, the decay of the top quark emitting a charged Higgs boson has the following expression:
\begin{eqnarray}
    \Gamma(t \to H^\pm b) = \frac{G_F m_t}{8 \sqrt{2} \pi}\bigg(m_t^2 |Y_{33}|^2+m_b^2|X_{33}|^2\bigg) \bigg(1-\frac{m_{H^\pm}^2}{m_t^2}\bigg)^2  \ .
\end{eqnarray}

\begin{center}
    \begin{figure}[ht]
    \centering
    \includegraphics[width=0.75\linewidth]{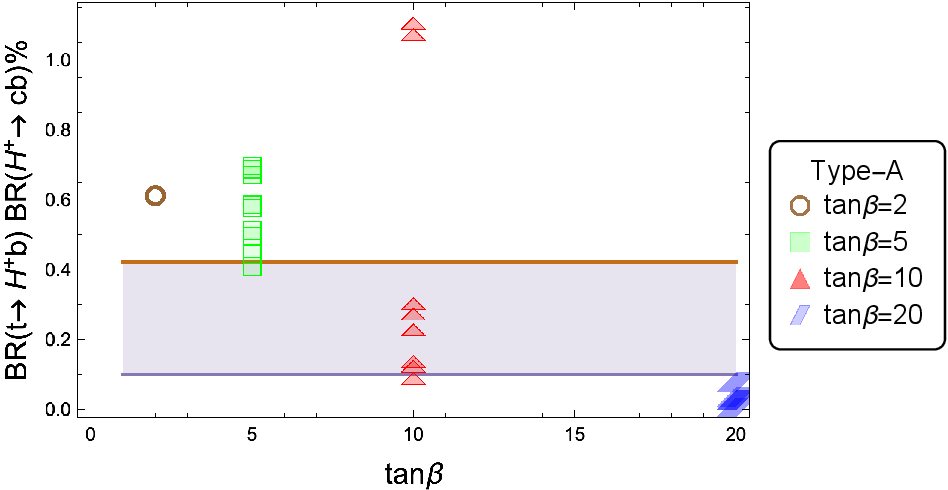}
    \vspace{1cm}
    \\
    \includegraphics[width=0.75\linewidth]{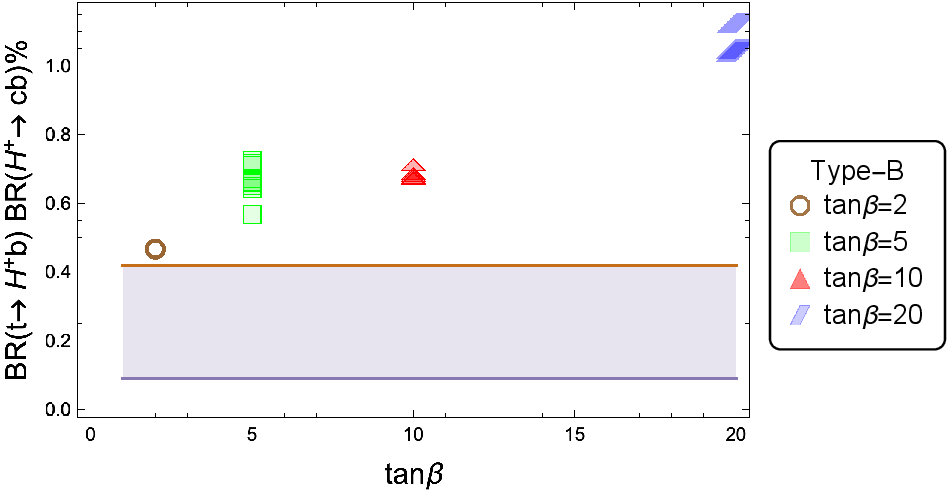}
    \caption{$BR(t\to H^\pm b)BR( H^\pm \to c b) \% $ vs. $\tan \beta$, benchmark points of the model are selected. The upper panel shows the THDMZ3 Type-A case while the THDMZ3 Type-B case in shown on the bottom panel. The shaded region is the allowed region for the experimental data of LHC \cite{ATLAS:2023bzb}.  }
    \label{fig:BRBR}
    \end{figure}
\end{center}

It is possible to determine the benchmarks points of the model using the points in Figure~\ref{fig:BRs} and imposing the ATLAS $95 \%$ confidence level exclusion bounds on the product of the branching fractions $BR(t\to H^\pm b) \times BR(H^\pm \to cb)$, reported as a function of $m_{H^\pm}$ in the range 60 GeV-160 GeV, which are between $0.15\%$ $(0.09)$ and $0.42\% $ $(0.25)$ from the observed (expected) limits, with center-of-mass energy $\sqrt{s}=13 $ TeV and integrated luminosity of $139 \ fb^{-1}$~\cite{ATLAS:2023bzb}, It is found that some points can satisfy the constraints only for the THDMZ3 Type-A, with $\tan \beta =5, 10 $ and $20$: one point for $\tan \beta =5$, one for $\tan \beta = 20$,  and five points for $\tan \beta = 10$ (as shown in Figure~\ref{fig:BRBR}).

\begin{center}
    \begin{figure}[ht]
    \centering
    \includegraphics[width=0.65\linewidth]{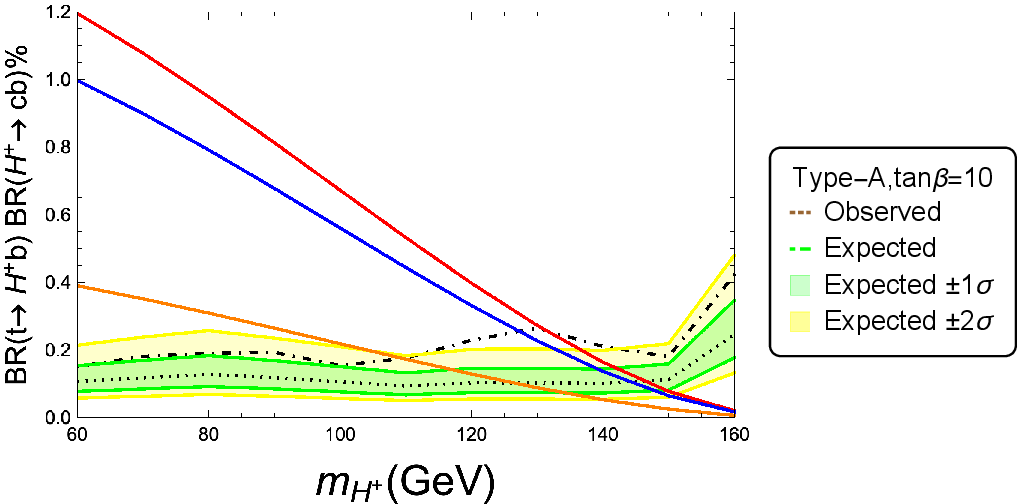}
    \caption{Contribution to $BR(t\to H^\pm b)BR( H^\pm \to c b) \% $ as a function of $m_{H^\pm}$ for three benchmark points of the model (solid lines in orange, blue and red). Only the two benchmark points corresponding to blue and red lines are in agreement with the slight excess for $m_{H^\pm}=130$ GeV reported by the ATLAS Collaboration~\cite{ATLAS:2023bzb}. }
    \label{fig:LHC}
    \end{figure}
\end{center}

In particular, focusing on the largest excess in the data for $m_{H^\pm}=130$~GeV reported by ATLAS~\cite{ATLAS:2023bzb} (with a global significance around $2.5 \sigma$ , with center-of-mass energy $\sqrt{s}=13 $ TeV and integrated luminosity of 139 $fb^{-1}$), only two points with $\tan \beta = 10$ can reproduce this slight excess, as shown in Figure~\ref{fig:LHC}. These benchmark points are taken as prospect for discovering a light charged Higgs boson in the future Large Hadron electron Collider (LHeC), where this signal could be studied in a complementary way~\cite{LHeC:2020van} (this could also happen at the Future Circular Collider, operating in a hadron-electron collision mode (FCC-he), which is foreseen as an improved proposal of the LHeC at an advanced stage of its development~\cite{FCC:2018byv}).    

Recapitulating: the benchmark points satisfy the bounds summarized in Table~\ref{tab:higgs}, Figure~\ref{fig:Bdn}, and Figure~\ref{fig:BRBR}, with a mass spectrum of $m_{h^0}=125$ GeV, $m_{A^0}>m_{H^\pm}$, 160 GeV$ \leq m_{H^0} \leq 260$ GeV, and 80 GeV $\leq m_{H^\pm} \leq 160$. The benchmark points are given in Table~\ref{tab:XY} in terms of the values for the $H^\pm cb$, $H^\pm cs$, and $H^\pm tb$ couplings.
\begin{center}
\begin{table}
    \centering
    \begin{tabular}{|c|c|c|c|c|c|c|}
    \hline  
      \cellcolor{gray!25} \ BPs \  & \ \ \cellcolor{gray!25} $X_{22}$ \ \ & \ \ \cellcolor{gray!25} $X_{23}$\ \  & \ \ \cellcolor{gray!25} $X_{33}$ \ \  & \ \ \cellcolor{gray!25} $Y_{23}$ \ \  & \ \  \cellcolor{gray!25}$Y_{22}$ \ \  & \ \ \cellcolor{gray!25} $Y_{33}$ \ \  \\
       \hline \hline
       BP-I & 187.23    & 4.33      & 5.33    &  0.087   & 0.43    &  0.099     \\
       \hline
       BP-II & 126.23    & 3.39      & 5.74    &  0.019   & 0.15    &  0.099   \\
      \hline
    \end{tabular}
    \caption{Values of the $H^\pm cb$, $H^\pm cs$ and $H^\pm tb$ couplings given in  \eqref{couplings-typeI} for the benchmark points of the model shown in Figure~\ref{fig:LHC} for THDMZ3 Type-A with $\tan \beta =10$.}
    \label{tab:XY}
\end{table}
\end{center}

In order to track down the relation of the $Z_3$ symmetry to the observed numerical enhancement, one can see the textures obtained for the ${\cal Y}^{qi}$ in Eqs. (9) and (10) that lead to the specific $\tan\beta$ dependence in the mass matrices (Eq. (12)).
 The enhancement behavior can be illustrated for the specific case of the $H^{\pm}$ coupling to charm and bottom quarks (see Eq.~20), where the $m_b\,X_{23}$ and $m_c\,Y_{23}$ couplings are given by
\begin{equation}
m_b\,X_{23} \sim (V_{CKM})_{23} \left[ X\,m_b - \frac{v\, f(X)}{\sqrt{2}}\,{\cal \hat{Y}}_{33}^{d1} \right] - \frac{v\, f(X)}{\sqrt{2}} \left[(V_{CKM})_{22}\, {\cal \hat{Y}}_{23}^{d1} + (V_{CKM})_{21}\, {\cal \hat{Y}}_{13}^{d1} \right],
\end{equation}

\begin{equation}
m_c\,Y_{23} \sim  \left[ Y\,m_c - \frac{v\, f(Y)}{\sqrt{2}}\,{\cal \hat{Y}}_{22}^{u1} \right]\,(V_{CKM})_{23} - \frac{v\, f(Y)}{\sqrt{2}} \left[{\cal \hat{Y}}_{21}^{u1}\,(V_{CKM})_{13}\, + {\cal \hat{Y}}_{23}^{u1}\,(V_{CKM})_{33} \right].
\end{equation}

The semi-analytic form of the coupling for THDMZ3 Type-A (for our BPs) is given by
 
 \begin{eqnarray}\nonumber
     m_b X_{23}&\sim& m_b \left[ z_1 \cot \beta+ z_2 \frac{\sqrt{1+ \tan\beta^2}}{\tan\beta} \frac{v}{m_b}\right] \\ &=& m_b \left[ 0.041 \cot\beta - (0.042- 031 i) \frac{\sqrt{1+ \tan\beta^2}}{\tan\beta} \frac{v}{m_b}\right]\\ \nonumber \\ \nonumber
     m_c Y_{23}&\sim& m_c \left[ z_3 \cot \beta + z_4 \frac{\sqrt{1+ \tan\beta^2}}{\tan\beta} \frac{v}{m_c}\right] \\ &=& m_c\left[ 0.041 \cot\beta - (0.032- 025 i) \frac{\sqrt{1+ \tan\beta^2}}{\tan\beta} \frac{v}{m_c}\right] \ , 
 \end{eqnarray}
where the $z_i$ are complex numbers parametrizing relations involving numerical values of the free parameters, obtained from the fit to quark masses and $V_{CKM}$ elements.

\section{Prospects for observing a light charged Higgs Boson}
\label{sec:prospects}

 As mentioned in the Introduction, one of the goals the LHeC will strive to achieve is to produce a cleaner signal of both charged and neutral Higgs bosons, due to several advantages, namely a reduction of the QCD background in hadron-hadron collisions, low pile up, simplification of final state topologies, and an improvement of the kinematical reconstruction of observables that involve Higgs-fermion interactions. As such, the LHeC could be considered a Higgs boson factory~\cite{LHeC:2020van}. In particular, the study of the production of a light charged Higgs boson followed by any mode decay in its final state, which would be an undeniable signal of new physics, can be considered one of its aims.

 The production of charged Higgs boson in the model presented in this letter can be analyzed by means of the process $e^- p \to H^- \nu_e q$, with $q=q_l$ (light quarks) or $q=b$, where $q_l=u, d, c, s $, followed by the decay channel $H^- \to b\bar{c}$ in the final state, with $BR(H^- \to b\bar{c})\sim 70\% $, $\tan \beta = 10$ and $m_{H^\pm}=130$ GeV. Following the analysis presented in~\cite{Flores-Sanchez:2018dsr} with the benchmark points in Figure~\ref{fig:LHC}, the $e^- p \to H^- \nu_e q$ cross-section can be computed for this process including the final state for the $H^-$ (relevant diagrams and background signals ($\nu_e jjj$, $\nu_e jjb$, $\nu_e jbb$, $\nu_e \nu_l l j $, $\nu_e \nu_l l$ and $\nu_e tb$ ) can be found in~\cite{Flores-Sanchez:2018dsr}) considering the LHeC with a center-of-mass energy $\sqrt{s} \approx 1.3 TeV $ and initial integrated luminosity of $L=100 fb^{-1}$, as well as foreseen luminosities of $1000 fb^{-1}$ till $3000 fb^{-1}$ for its last stages. 
 
\begin{center}
    \begin{figure}[ht]
    \centering
    \includegraphics[width=0.5\linewidth]{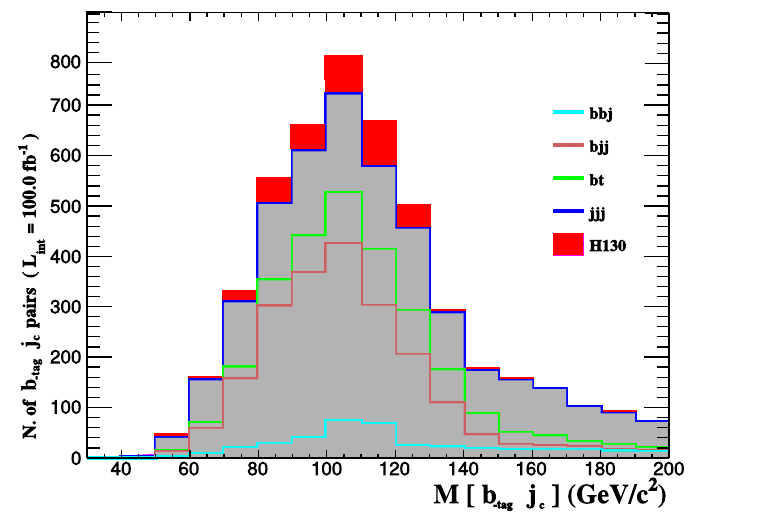}
    \caption{Distributions for $(m_{H^\pm}-20 \ {\rm GeV})$ $< M((b_{tag}, j_c))< m_{H^\pm}$, where $ M((b_{tag}, j_c))$ is the invariant mass of two central jets for $m_{H^\pm}= 130$ GeV.}
    \label{fig:Histo}
    \end{figure}
\end{center}

 The most relevant process for this work is $e^- p \to H^- \nu_e b$ followed by $H^- \to b\bar{c}$ (note that the contribution of light quarks are also considered in the simulation). In general one finds $3 j + \cancel{E}_T $ in the final state, where $j$ is a generic jet  and $\cancel{E}_T $ is the missing transverse energy. In the reconstruction of the charged Higgs there is a light jet, an associated $b$-tagged one, and another jet which could be $b$-tagged. As done in~\cite{Flores-Sanchez:2018dsr}, the numerical analysis implements: MAdGraph as a parton-level generator~\cite{Alwall:2014hca}, where Pythia8 is included as parton shower, hadronization and hadron decays~\cite{Bierlich:2022pfr}, Delphes as an emulator detector~\cite{deFavereau:2013fsa}, and Madnalysis5 for Monte Carlo event generators~\cite{Conte:2012fm}. The following selection/rejection of the signal were implemented:
 
\begin{itemize}
    \item Selection I: a signal with at least one b-tagged jet and with an efficiency of $12 \%$, while the backgrounds $\nu_e bbj$, $\nu_e jjb$,  $\nu_e bt$  and $\nu_e jjj$,  have an efficiency of $10 \% $, $8 \%$, $5 \%$ and $1 \%$, respectively.
    \item Selection II: two central jets in the detector, one b-tagged and one with a light quark labeled as $j_c$, with $P_T (b_{tag}) > 30$~GeV and $P_T (j_c) > 20$~GeV, where $P_T$ is the transverse momentum, followed by a cut on the pseudorapidity $|\eta (b_{tag}, j_c)| > 2.5$. Considering the standard cone separation    $1.8 <\Delta R (b_{tag} j_c) <3.4$, one can get the cumulative efficiency of the signal to be $7.3 \% $  while the backgrounds $\nu_e bjj$, $\nu_e jbb$,  $\nu_e bt$  and $\nu_e jjj$,  have a cumulative efficiency of $6 \% $, $3.7 \%$, $3.3 \%$ and $0.3 \%$, respectively.
    \item Selection III: a third generic jet  with $|\eta| > 0.6  $ and $P_T (j) >20$~GeV. With this assumption, the signal has an efficiency of $5.4 \%$, while the best efficiency for the background is for the signal  $\nu_e bjj$, and for the other background signals $\nu_e jbb$,  $\nu_e bt$  ($\nu_e jjj$) the efficiencies are below $2\%$ ($0.3 \%$). 
    \item Selection IV: taking in account the two central jets preselected $b_{tag}$ and $j_c$, one can get events in the invariant mass of the aforementioned jets associated with the signal for $m_{H^\pm}=130$~GeV. Considering that at detector, one may suffer a mass shift due to jets dynamics, one can establish the following selection: $(m_{H^\pm}-20 \ {\rm GeV})$ $< M((b_{tag}, j_c))< m_{H^\pm}$, rejecting the invariant mass of light central jets, which are associated with a hadronic $W^\pm$ boson decay (see~\cite{Flores-Sanchez:2018dsr}). In Figure~\ref{fig:Histo} one can see the distributions of invariant mass  $ M((b_{tag}, j_c))$ of the aforementioned pair of central jets and compare them to the corresponding background spectra. The previous cuts prioritize the signal, providing anefficiency of $2.4\%$, while the background signals at $0.6 \%$ are reduced.
\end{itemize}

Table~\ref{tab:Events} lists the cross-section, branching ratios, and events rates at parton level for the two benchmark points of the model. The  selection procedure above led to the results in Table~\ref{tab:SB}, where each selection step (I - IV) is shown for each benchmark point, as well as their respective ratios between signal significance and cumulative backgrounds $S/\sqrt{B}$, with promising values of $4.8$ and $7.04$ (for an initial integrated luminosity  of $L= 100 fb^{-1}$). Overall the improvement of the  $H^\pm cb$ and $H^\pm tb$ couplings can be traced down to  the role of the $Z_3$ flavored doublet and quark sector.
\begin{center}
\begin{table}[ht]
  \centering
    \begin{tabular}{|c|c|c|c|c|c|c|}
    \hline 
\cellcolor{gray!25} BP  & \cellcolor{gray!25} \ $m_{H^\pm}$ \ & \cellcolor{gray!25} \ $\sigma (e^- p \to \nu_e H^- q)$ (pb) \ & \cellcolor{gray!25} \ $BR(H^- \to b \bar{c}) \% $ \  & \cellcolor{gray!25} \  $\sigma \times BR \times L $ \ \\
       \hline \hline
\ BP-I \  & \ 130 GeV \      & $1.124 \times 10^{-1} $               & $ 68 \% $                   &  7643 \\
       \hline
\ BP-II \ & \ 130 GeV \      & $1.824 \times 10^{-1}$                &  $61 \% $                   &  11126 \\
      \hline 
    \end{tabular}
    \caption{Cross-section, branching ratios, and event rates at parton level for THDMZ3 Type-A with $\tan \beta =10$ and an integrated luminosity $L= 100 fb^{-1}$ for Benchmark Points (BP) of the model.}
    \label{tab:Events}
\end{table}
\end{center}

\begin{table}[ht]
    \centering
    \begin{tabular}{|c|c|c|c|c|c|c|}
    \hline 
\cellcolor{gray!25} BP  & \cellcolor{gray!25} Event (raw)  & \cellcolor{gray!25}  Selection I   & \cellcolor{gray!25}  Selection II  &  \cellcolor{gray!25} Selection III  & \cellcolor{gray!25}  Selection IV  & \cellcolor{gray!25} \ $\mathbf{S/\sqrt{B}}$ \ \\
       \hline \hline
\ BP-I \ &  7643        & 917           &  558          &  412 & 183  &  {\bf 4.8} \\
       \hline
\ BP-II \ &  11126      & 1134     &  812   & 600     & 266 & {\bf 7.04} \\
      \hline
    \end{tabular}
    \caption{Raw and selected events (at each step as described in the text) and signal to background ratios for the two benchmark points of the THDMZ3 Type-A model with $\tan \beta =10$ and an integrated luminosity $L= 100 fb^{-1}$. }
    \label{tab:SB}
\end{table}
\section{Conclusion}
\label{sec:conclusion}
This letter considers a Two Higgs Doublet Model whose couplings to SM fermions involve a $Z_3$ flavor symmetry that can lead to enhancements for the charged Higgs coupling to $cb$, in 
such a way that it may be possible to produce and detect them at the future LHeC. The study includes a determination of the free parameters involved in the quark mass matrices and the quark mixing $V_{CKM}$ matrix performed using a Particle Swarm Optimization algorithm, followed by the flavor and Higgs physics constraints coming from meson decays, $b \to s\gamma$, $B_0 - \overline{B}_0$ mixing, and the neutron's electric dipole moment. By imposing the current exclusion bounds from ATLAS, it was shown that several benchmark points satisfy them and two of them are in agreement with the ($2.5 \sigma$ global significance) branching ratio excess observed by ATLAS for $m_{H^{\pm}}=130$~GeV. Charged Higgs production and signal to background ratios were computed for these benchmark points assuming an initial integrated luminosity of $L=100 fb^{-1}$ at the LHeC. Promising values were obtained for both.

\section{Acknowledgments}
Dedicated to the memory of P.Q. Hung. AA, JHS, and RNP acknowledge support from SNII-SECITHI. 
AM acknowledges support from  SECIHTI - graduate fellowship 2145260. 
JHS is also supported by VIEP-BUAP, PRODEP (Mexico) under grant: "Higgs and Dark matter Physics" (SECIHTI -CBF-2025-G-1187), and by SECITHI under a sabbatical 2025 grant.

\bibliographystyle{ieeetr}

\end{document}